\documentclass[journal]{IEEEtran}
\usepackage{amsfonts,subfigure,multicol,color,verbatim,
graphicx,cite,epsfig,amssymb,amsmath,cases,bm,algorithm,
algorithmic,xcolor,multirow}

\begin{document}
\title{A Data Mining Approach Combining K-Means Clustering with Bagging Neural Network for Short-term Wind Power Forecasting}

\author{\IEEEauthorblockN{Wenbin~Wu and Mugen~Peng}
\thanks{Wenbin~Wu (e-mail: binge@bupt.edu.cn) and Mugen~Peng (e-mail: pmg@bupt.edu.cn) are with the Key Laboratory of Universal Wireless Communications (Ministry of Education), Beijing University of Posts and Telecommunications, Beijing, China.}}

\maketitle

\begin{abstract}

Wind power forecasting (WPF) is significant to guide the dispatching of grid and the production planning of wind farm effectively. The intermittency and volatility of wind leading to the diversity of the training samples have a major impact on the forecasting accuracy. In this paper, to deal with the training samples dynamics and improve the forecasting accuracy, a data mining approach consisting of K-means clustering and bagging neural network is proposed for short-term WPF. Based on the similarity among historical days, K-means clustering is used to classify the samples into several categories, which contain the information of meteorological conditions and historical power data. In order to overcome the over fitting and instability problems of conventional networks, a bagging-based ensemble approach is integrated into the back propagation neural network. To confirm the effectiveness, the proposed data mining approach is examined on real wind generation data traces. The simulation results show that it can obtain better forecasting accuracy than other baseline and existed short-term WPF approaches.

\end{abstract}
\begin{IEEEkeywords}
\centering Wind power forecasting, k-means clustering, ensemble learning, bagging neural network
\end{IEEEkeywords}

\section{Introduction}

The development and utilization of renewable energy has been one of the hottest spots around the world. Wind power generation is rapidly expanding into a large-scale industry due to the cleanness and wide availability, and has been characterized as an fluctuating and intermittent power. Unfortunately, it is difficult to ensure the security and stability while accessing to the electricity grid, especially for large-scale application. An accurate and reliable wind power forecasting approach is essential for power quality, reliability management while reducing the cost of supplying spinning reserve.

A large number of wind power forecasting approaches have been proposed in many literatures, which can be roughly classified into three categories: 1) physical forecasting approach; 2) statistical forecasting approach; 3) combination approach \cite{bib:1}. The physical approach uses detailed topological and meteorological descriptions to model the conditions at the location of the wind farm \cite{bib:2}. Then the wind speed is predicted by numerical weather forecasting approach and converted into wind power via the power curves generated by wind turbines. The accuracy of this approach largely depends on the amount of physical information. The statistical approach aims at establishing the relationship between wind power and a set of variables including historical data and online measured data such as wind speed and wind direction \cite{bib:3}. This approach is applicable for most scenarios without considering geographical conditions. The core idea of the combination approach is to take advantage of the physical and statistical methods and improve the forecast accuracy \cite{bib:4}.

The physical approach can establish a specialized wind power forecasting model for a wind farm without large amount of historical data. In particular, this approach needs detailed physical characteristics of wind turbines and wind farm to achieve an accurate model \cite{bib:5}. However, it is difficult to collect these physical characteristics in a short period of time. Besides, the model has poor versatility due to the factors of specific geographical conditions and operation state of wind turbines.

Compared with the physical approach, the statistical approach needs to collect a lot of historical data to build the forecasting model. In conventional statistical approaches, the time series model is usually applied to predict the wind power depending on the principle of continuity. There are several kinds of time series models, including autoregressive model (AR), moving average model (MA), autoregressive moving average model (ARMA), and autoregressive integrated moving average model (ARIMA) \cite{bib:9}-\cite{bib:12}. Reference \cite{bib:13} concentrates on multi-time series and multi-time scale modeling in wind speed and wind power forecasting. Then AR, MA, ARMA and ARIMA approaches are utilized to model multi-time series. The Kalman filter approach \cite{bib:14} is applied to optimize the time series model, which can modify the weights dynamically on the basis of ARMA model. Much work have shown that the time series models can obtain good forecasting accuracy only in the very short-term scale scenarios. In addition, this approach is applicable for the continuous sequence without considering the similarity among historical data.

Data mining approaches are catching researcher's attention \cite{bib:6}-\cite{bib:8}. It has been reported that neural network (NN) outperformed other methods in short-term forecasting problems \cite{bib:19}. NN is able to model the complex nonlinear relationship between the historical data and forecasting power \cite{bib:15}. Besides, NN needs to use the historical data which have a major impact on the wind power as the input variables, such as wind speed and temperature. Back propagation (BP) NN is easy to formulate and often used as a reference model. But there are some potential improvements for the BPNN. Firstly, the result of the BPNN is sensitive to the initialization of weights and biases, and the network is easy to fall into local optimum. Secondly, the overtraining problem may arise if the BPNN has too many parameters to be estimated in the training samples \cite{bib:16}. Finally, several improvements for the BPNN are achieved by 1): optimizing parameters 2): tuning the structure of NN 3): changing the kernel of NN. Particle swarm optimization (PSO) \cite{bib:20}, the Elman NN \cite{bib:17} and Radial Kernel Function (RBF) as examples respectively correspond to the above approaches. However, these improvements don't consider that the selection of training samples has an important impact on the forecasting accuracy.

In order to improve the forecasting accuracy, clustering approaches are also applied to select proper training set. Reference \cite{bib:17} proposes an approach which selects the similar days as the training samples. The K-means clustering is utilized to classify wind speed into several categories. In reference \cite{bib:21}, the ARMA model is presented and the K-means clustering is used to classify wind direction. However, only the weather conditions are clustered and the actual operation state of wind turbines is not considered.

In this paper, a data mining approach consisting of the K-means clustering and bagging neural network is proposed to predict the short-term wind power of individual wind turbine. The main contributions of this paper are summarized as follows.
\begin{enumerate}
\item To mitigate the impact of the diversity of training samples, the K-means clustering is utilized to classify the original data sets into several categories according to meteorological conditions and historical power. Person correlation coefficient is used to calculate the similarity between each category and the forecasting day. The most similar category is selected as the training sample.
\item To overcome the instability and over fitting problems of the BPNN, a bagging-based approach is integrated into the BPNN as the forecasting engine. The random sampling is conducted on the training sample by the Bootstrap sampling technique to form ${N}$ subsets. The BP algorithm is applied to each subset and trained on the individual network. The final result is determined by calculating the average value of ${N}$ networks.
\item The performance of the proposed approach is evaluated. Simulation results show that the root mean squared error (RMSE) and mean absolute error (MAE) of the proposed approach are less than other baseline approaches, which demonstrates the proposed approach can improve the forecasting accuracy.
\end{enumerate}

Besides, many researchers have focued on the platform of collecting and processing data in wind turbines because they are important for managing and forecasting the wind energy. In order to make use of the manifold potential, the available data of wind turbines needs to be analyzed and the advanced measuring technologies and communication networks should be implemented. The Cyber Physical System (CPS) is a powerful approach to develop the wind energy sector, which has been widely used in CMS or SCADA systems \cite{bib:24}.

The rest of the paper is organized as follow. Section \uppercase\expandafter{\romannumeral2} presents the process of establishing the prediction model. Section \uppercase\expandafter{\romannumeral3} carries out the results of the simulation and Section \uppercase\expandafter{\romannumeral4} concludes the paper and summarizes the future work.

\section{The establishment of the wind power forecasting model}

Data mining approaches have been widely used for classification and prediction problems. The proposed approach is based on data mining, which consists of the K-means clustering and bagging neural network. Fig. 1 shows the wind power forecasting model. Firstly, data preprocessing is conducted on the vector space to clean unreasonable data, normalize the training samples and select the most related variables as the inputs of the neural network. Secondly, data after preprocessing are clustered by the K-means clustering to select the training set which is most similar to the forecasting day. Finally, the wind power is forecasted by the bagging neural network, which is able to alleviate the instability and over fitting problems of the BPNN.
\begin{figure}
\centering \vspace*{0pt}
\includegraphics[scale=0.5]{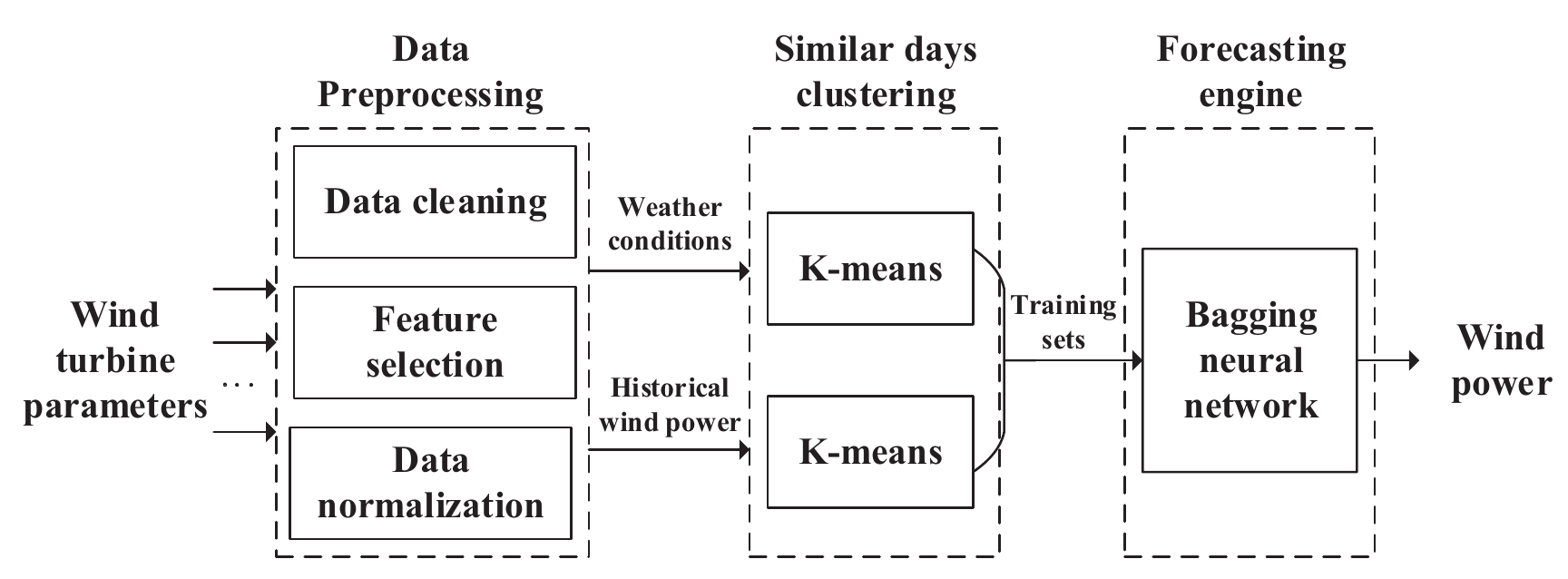}
\setlength{\belowcaptionskip}{-100pt} \caption{\textbf{The system
architecture for the proposed WPF approach}} \label{HRAN}\vspace*{-10pt}
\end{figure}

\subsection{Data preprocessing}
A number of wind turbine parameters are collected as the training samples via the sensor unit. However, these samples may contain unreasonable data. Besides, using too many parameters as the training features would increase the computing complexity and obtain undesired results for the reason that some variables are irrelevant or redundant in this model. Selecting features which are most related to the wind power is able to improve the accuracy. Finally, data normalization has an effect on the convergence rate and accuracy of the training algorithm. Thus, in order to obtain accurate forecasting results, data preprocessing is necessary.

\begin{itemize}
 \item \emph{Data cleaning}: The original samples may contain data whose values of some characteristics are unreasonable. For example, the values of wind speed and wind power are less than zero. It is obvious that these data are not available and need to be removed. Then, it is necessary to fill the vacancy of the deleted data. The mean value method is applied, which makes use of the mean value ahead and back of the deleted data.
 \item \emph{Feature selection}: In theory, more input variables can carry more discriminating power. But in practice, excessive variables are prone to cause many problems. Therefore, selecting a suitable set of input variables from the raw data has a great impact on the forecasting performance. Relief algorithm is a kind of feature weighting algorithms. The core idea is that the different weights are assigned to the corresponding features according to the correlation, and the feature whose weight is less than the threshold would be removed. The formulation of Relief algorithm is given by \cite{bib:18}. The running time of Relief algorithm increases linearly with the sampling times and original features so that this method has high operating efficiency. In addition, Relief algorithm can achieve the purpose of physical dimensions reduction compared with the principal component analysis (PCA).
 \item \emph{Data normalization}: The aim of data normalization is to transform the raw data to the same orders of magnitude so that the convergence rate and forecasting accuracy can be improved. The min-max method is applied for normalization, which can be expressed as ${\overline{x}=(x-min)/(max-min)}$. In this equation, x is the original data, and max and min represent the maximum and minimum value of the training set. The result ${\overline{x}}$ is mapped to [0,1].
\end{itemize}

The three steps above are significant to more accurate forecasting results. After data preprocessing, the proposed approach can be implemented and compared with other WPF approaches.

\subsection{Similar day clustering based on meteorological conditions and historical power}
Large amount of data are generated by wind turbines when operating normally. The training samples can be regarded as the time series and auto regressive method is often utilized to build the forecasting model. However, the random behavior of wind can lead to the inconsistencies of training samples. Then, it is difficult to obtain desire results when using this kind of training samples. Therefore, selecting similar days and classifying them into the same category as the training samples can be considered as a good method to improve the forecasting accuracy.

\begin{enumerate}
\item \textbf{\emph{The clustering sample}} \\
Wind speed and temperature have a major impact on wind power generation. Selecting the samples which have the most similar wind speed and temperature with the forecasting day from the historical data is significant. A sample consisting of wind speed and temperature is used as the clustering benchmark, which is constructed as
\begin{equation}
 S_{1} = [{W{S_{\max }},W{S_{\min }},W{S_{{\mathop{\rm mean}\nolimits} }},{T_{\max }},{T_{\min }},{T_{{\mathop{\rm mean}\nolimits} }}}],
\end{equation}
where ${WS_{max}}$, ${WS_{min}}$, ${WS_{mean}}$, ${T_{max}}$, ${T_{min}}$ and ${T_{mean}}$ represents the maximum, minimum and average values of wind speed and temperature respectively.

According to the investigation results, wind power generation is largely determined by the meteorological conditions. However, the similar meteorological conditions can not represent that the wind power of the training samples is similar to the same of the forecasting day because of the different geographical conditions and the operation state of wind turbines. Therefore, the historical data which have the similar wind power with the forecasting day is also used for clustering. Then the sample ${S_{2}}$ could be expressed as
\begin{equation}
 S_{2} = [WP_{max},WP_{min},WP_{mean}],
\end{equation}
where ${WP_{max}}$, ${WP_{min}}$, ${WP_{mean}}$ represents the maximum, minimum and average values of daily wind power generation. Although the ${S_{2}}$ of the forecasting day is unknown, the day which is before the forecasting day can be assumed to have the same ${S_{2}}$. According to the principle of wind power generation, the daily wind power generation curve has the same trend for next few days in the same season. Then the samples ${S_{1}}$ and ${S_{2}}$ are clustered respectively. It is assumed that cluster 1 from ${S_{1}}$ and cluster 2 from ${S_{2}}$ are most similar to the forecasting day. Finally, the same days from cluster 1 and cluster 2 are collected as the training samples of the forecasting model.

\item \textbf{\emph{The clustering algorithm}} \\
Clustering belongs to unsupervised learning, which has no label on the training data. In this paper, it is needed to use the clustering algorithm to mine the similar samples and category them into one class. Typical clustering algorithms contain the K-means clustering and expectation maximization (EM).

In this paper, the K-means clustering is used as the clustering method for the reason that K-means algorithm can handle large amount of data sets effectively. K-means algorithm is a kind of clustering method based on partitioning, which usually judges the similarity by computing the distance. The core of K-means algorithm is to choose k center points randomly and partition the data according to the distance between the data and k center points. By means to the Euclidean distance, the algorithm assigns each data to its closest center point ${P_{k}}$, which is calculated by
\begin{equation}
P_{k} = \frac{1}{{{N_{k}}}} \cdot \sum\limits_{i = 1}^{{N_{k}}} {x_{i}^k},
\end{equation}
where ${x_{i}^k}$ is the i-th data in the cluster k, and ${N_{k}}$ is the number of data points in the respective cluster. The clustering center needs to be updated by computing the average value of every cluster until the value doesn't change any more. Then the initial data set is divided into k clusters in which the data have high similarity. The process of clustering is as follows:

\begin{itemize}
 \item \emph{Step 1}: K objects are selected as the initial clustering centers from the data set which obtains N objects.
 \item \emph{Step 2}: The distance between the data objects and clustering center is calculated and classifying them to the nearest cluster.
 \item \emph{Step 3}: The mean value of the cluster is computed and used to update the clustering center.
 \item \emph{Step 4}: The iterative method is applied in Step 2 and 3 until the value of the clustering center changes no more. If not, the process continues.
\end{itemize}

The K-means clustering is conducted on the data sets of ${S_{1}}$ and ${S_{2}}$. The data series are cut up into M components, and the training set can be expressed as ${D={d_{1},d_{2},...,d_{M}}}$, where D represents the training days. The K-means algorithm can select similar days from D as one class. Then several classes are produced. According to Pearson correlation coefficient, the most related class is chose as the final input of the neural network.
\end{enumerate}

\subsection{Bagging neural network}
Neural network (NN) has been one of the most effective data mining approaches for prediction. NN can deal with nonlinear problems well without establishing complex mathematical model. Back propagation neural network (BPNN) as one of the most common NNs is usually used as the forecasting algorithm. The basic BPNN consists of three layers: an input layer, a hidden layer and an output layer.
\begin{figure}
\centering \vspace*{0pt}
\includegraphics[scale=0.6]{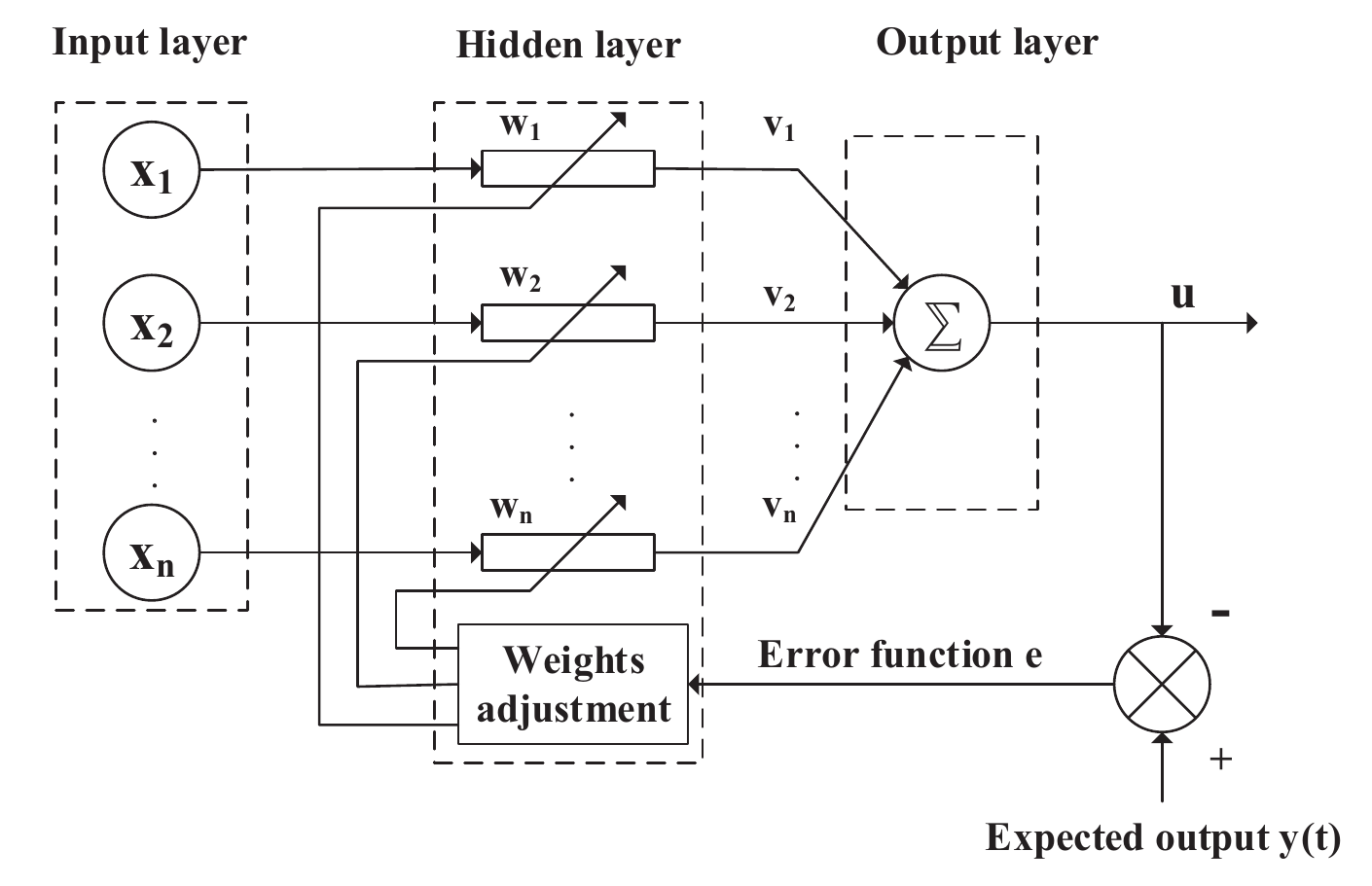}
\setlength{\belowcaptionskip}{-100pt} \caption{\textbf{The model of three-layers BPNN}} \label{HRAN}\vspace*{-10pt}
\end{figure}

Fig. 2 shows the principle of the BPNN. The BPNN consists of two processes: forward propagation of data stream and back propagation of the error signal. In the process of forward propagation, the state of neurons in each layer only affect ones in the next layer. If the expected output could't be obtained in the output layer, the algorithm turns to the process of back propagation of the error signal. The gradient descent method is conducted on the weights vector space. It is needed to dynamically search for a set of weights vector and minimize the error function. As for the hidden layer, the neural numbers of this layer is usually ${2M+1}$ according to the experience, where ${M}$ represents the neural numbers of the input layer. However, different neural numbers have an effect on the results of the output layer so that the network is tested when ${M = [2,3,...,10]}$ to find the best forecasting result.

In Fig. 2, ${[{x_{1},x_{2},...,x_{n}}]}$ are the input variables, ${[{w_{1},w_{2},...,w_{n}}]}$ are weights between the input layer and the hidden layer and ${[{v_{1},v_{2},...,v_{n}}]}$ are weights between the hidden layer and the output layer. ${Y(t)}$ is the expected output. In addition, the general Sigmoid function is used as the transfer function to handle the nonlinear problem, which can be expressed as
\begin{equation}
y = f\left( x \right) = \frac{1}{{1 + {e^{ - x}}}}.
\end{equation}

The output of the hidden layer ${z}$ can be formulated as
\begin{equation}
z = {f_{1}}\left( {\sum {{w_{i}}{x_{i}}} } \right).
\end{equation}

Similarly, the output of the out layer can be formulated as
\begin{equation}
u = {f_{2}}\left( {\sum {{v_{k}}{z_{k}}} } \right).
\end{equation}

Thus, the process of forward propagation is completed. Then, the error signal ${e}$ is produced by the function of ${u}$ and ${Y(t)}$, which is calculated as
\begin{equation}
e = \frac{1}{2}{\sum {\left( {{y_{i}}\left( t \right) - {u_{i}}} \right)} ^2}.
\end{equation}

The gradient descent method is utilized to modify ${w_{i}}$ and minimize the error function ${e}$. This method needs to calculate the partial derivatives to update the weighs of ${w_{i}}$ and ${v_{i}}$. It is incessantly required to be repeated until the value of ${e}$ equals to zero, which means that the values of ${u}$ and ${Y(t)}$ are identical. In this process, several weights need to be modified, which consumes a lot of time.

The BPNN can obtain good results compared with other algorithms such as linear aggression and support vector machine (SVM) when dealing with large amount of nonlinear samples. However, it is easy to trap into the local optimization and the forecasting result would be instable when the training samples change little. Besides, the over fitting problem would arise for the BPNN.

In order to improve the problems above, ensemble learning is applied to optimize the BPNN. Ensemble learning can combine several learning modules to enhance the stability and forecasting accuracy of the model. Bootstrap aggregating (bagging) algorithm as a kind of ensemble learning methods is used to improve the performance of the BPNN. It is verified that the bagging-based approach is effective for instable learning algorithms.

The bagging algorithm can be effective on the premise that a weak learner is applied to the training samples. Because of the poor accuracy of the weak learner, it is needed to use the learner several times on the training samples. Then a sequence of forecasting functions are produced and the most accurate function is determined by voting.

\begin{figure}
\centering \vspace*{0pt}
\includegraphics[scale=0.5]{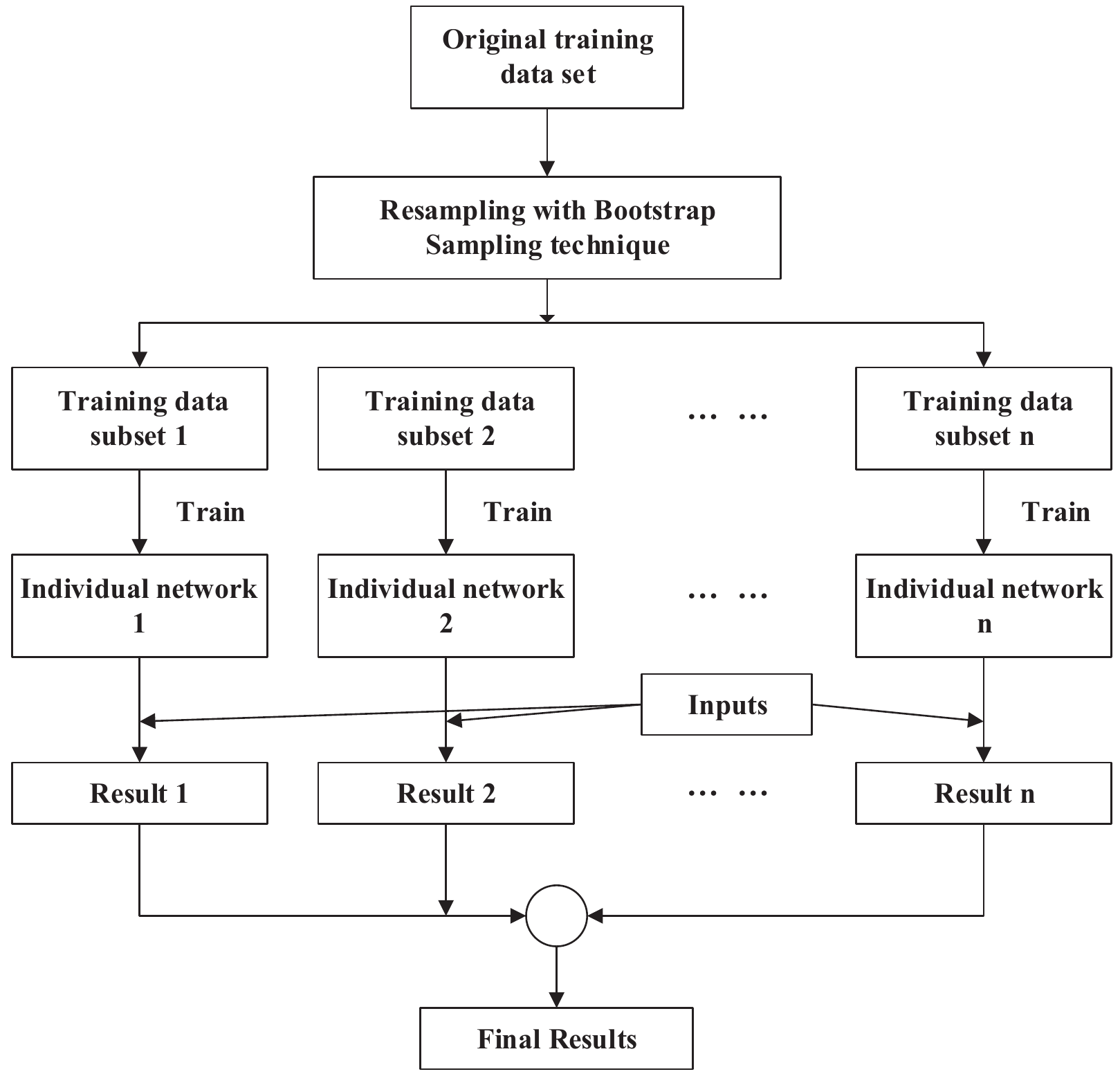}
\setlength{\belowcaptionskip}{-100pt} \caption{\textbf{The process of sampling and training of the bagging algorithm}} \label{HRAN}\vspace*{-10pt}
\end{figure}

Fig. 3 shows the process of the bagging algorithm. It can be seen that the original training data set is sampled again by Bootstrap sampling technique and the learning algorithm is applied for the subsets. Suppose ${T(x)}$ is a classifier, such as a tree, producing a predicted class label at input point ${x}$. To bag ${T}$, we draw bootstrap samples ${{\left\{ {\left( {x_{i}^*,y_{i}^*} \right)} \right\}^1},...,{\left\{ {\left( {x_{i}^*,y_{i}^*} \right)} \right\}^B}}$ each of size ${n}$ with replacement from the training data. Then the results can be expressed as follows:
\begin{equation}
{C_{bag}}\left( x \right)=Majority Vote\left\{ {{T^{*b}}\left( x \right)} \right\}_{b = 1}^B
\end{equation}
The output ${C_{bag}}\left( x \right)$ is the optimal classifier, which is determined by voting. In this paper, the bagging-based method is used for prediction. The key factor that bagging algorithm can improve the forecasting accuracy is the stability of the learning algorithm. Firstly, bagging algorithm can select K sets with n samples randomly from the initial training set. Secondly, BPNN is utilized to train the subsets for several times. Then a sequence consisting of the forecasting models ${{h_{1},h_{2},...,h_{K}}}$ is produced. Finally, the chosen forecasting model ${H}$ is determined by calculating the average value of ${K}$ models. In particular, all the forecasting models have the same importance.

Applying the BPNN based on the bagging algorithm can not only improve the forecasting accuracy but also reduce the time costs.

\section{Simulation results}
In this section, the proposed forecasting approach is simulated and compared to other approaches. The 10-min data obtained from the Supervisory Control And Data Acquisition system (SCADA) of a wind farm is adopted as research data. The six numbers for an hour are sum to obtain hourly data, which could avoid that 10-min data fluctuated greatly so that the forecasting accuracy could be improved.

In order to evaluate the performance, two kinds of error calculation methods: 1) root mean squared error (RMSE) and 2) mean absolute error (MAE) are proposed:

\begin{equation}
RMSE = \sqrt {\frac{1}{N}\sum\limits_{i = 1}^N {\mathop {\left( {{y_{i}} - y^{'}_{i}} \right)}\nolimits^2 } }
\end{equation}

\begin{equation}
MAE = \frac{1}{N}\sum\limits_{i = 1}^N {\left| {{y_{i}} - y^{'}_{i}} \right|}
\end{equation}
In the above equations, ${y_{i}}$ is the actual power, ${y^{'}_{i}}$ is the predictive power, N is the number of the test data.

\subsection{Data preparing}
In this paper, the measured data from September 29th to September 30th for a total of 24 hours are selected as test data. Data in three months before this period of time are used as training samples for clustering. Fig. 4 shows the wind power generation of all the year. It can be observed that the power generation is much different before and after July. Even in the same season, the power curve varies dramatically, which means that it is necessary to apply the clustering method.
\begin{figure}
\centering \vspace*{0pt}
\includegraphics[scale=0.23]{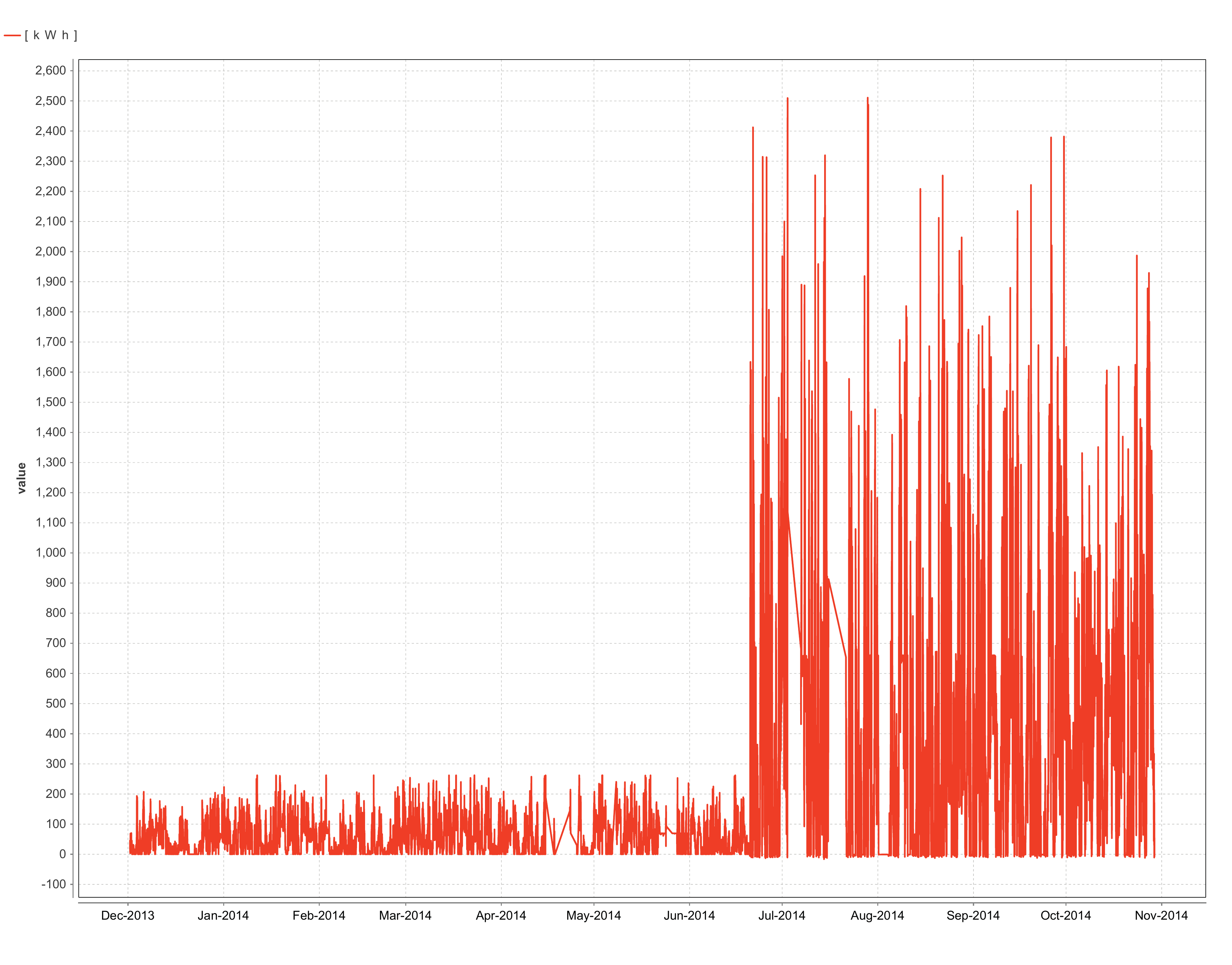}
\setlength{\belowcaptionskip}{-100pt} \caption{\textbf{The wind power generation curve from Dec. 2013 to Nov. 2014}} \label{HRAN}\vspace*{-10pt}
\end{figure}
There are 39 variables collected by the SCADA system. Relief algorithm is conducted on the vector space to select proper variables as the inputs of neural networks. The results of Relief algorithm have been shown in Table \uppercase\expandafter{\romannumeral1}, in which the variables whose weight is large than 0.01 are listed. According to the results, the first four variables have better weights. Besides, the ambient temperature is also significant. Then, the whole input variables are selected.
\begin{table}
\caption{The results of Relief algorithm}\label{survey}
\scriptsize 
\centering
\begin{tabular}{|p{2 in}|p{0.9 in}|}
 \hline
 \textbf{Variable} & \textbf{Weight} \\ \hline
 \multirow{1}{2 in}{Average wind speed per second}
 & 0.052 \\ \hline
 \multirow{1}{2 in}{Machine side semiconductor temperature}
 & 0.022 \\ \hline
 \multirow{1}{2 in}{Network side semiconductor temperature}
 & 0.016\\ \hline
 \multirow{1}{2 in}{The maximum angle of blade 1}
 & 0.013\\ \hline
 \multirow{1}{2 in}{The angle of blade 1}
 & 0.003\\ \hline
 \multirow{1}{2 in}{Generator bearing temperature}
 & 0.003\\ \hline
 \multirow{1}{2 in}{The ambient temperature}
 & 0.002\\ \hline
 \multirow{1}{2 in}{Rotor winding temperature of generator}
 & 0.002\\ \hline
\end{tabular}
\end{table}

Table \uppercase\expandafter{\romannumeral1} illustrates that wind speed is the most important variable in wind power forecasting. Although temperature from the wind turbine units plays an important role according to Relief algorithm, it can not be utilized as the forecasting variable for the reason that it is unable to obtain the temperature of machines in the forecasting day. The generator bearing temperature can not be used for the same reason. Therefore, we select average wind speed per second, the angle of blade 1 and the ambient temperature as the input variables, where the angle of blade represents the wind direction. Besides, the value of the angle of blade is decomposed into sine and cosine values.

Fig. 5 shows the relationship between wind speed and power. It can be seen that when actual wind speed is less than cut-in wind speed, and the wind turbines do not operate and the wind power is zero. When the actual wind speed is larger than cut-in wind speed, the value of wind power is increasing as the wind speed growing until the wind speed is less than cut-out wind speed. When the actual wind speed is larger than cut-out wind speed, wind turbines would operate in the state of rated power. It is obvious that the relationship between them is nonlinear so that applying the neural network is necessary.
\begin{figure}[h]
\centering \vspace*{0pt}
\includegraphics[scale=0.2]{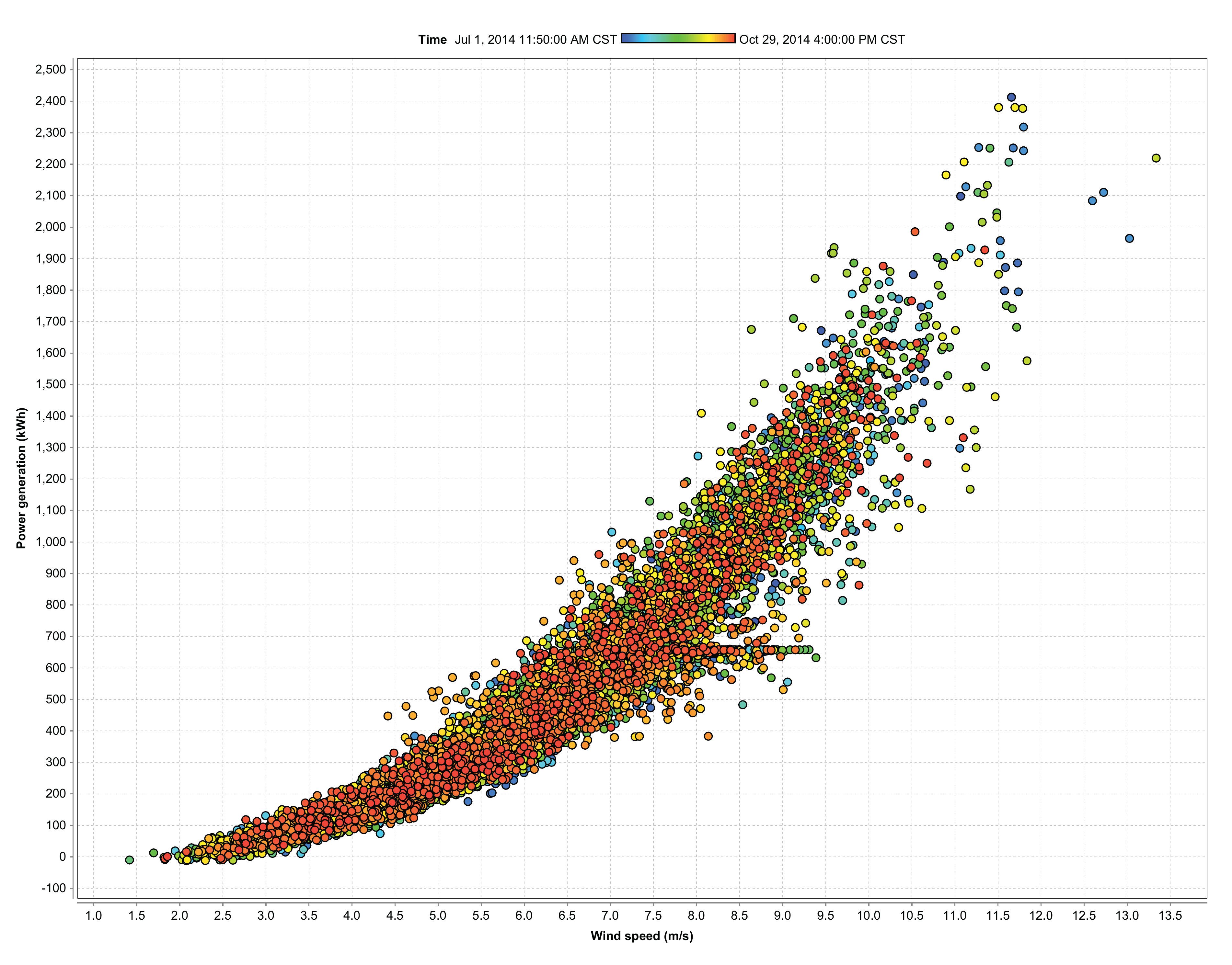}
\setlength{\belowcaptionskip}{-100pt} \caption{\textbf{Strong correlation between wind power generation and wind speed}} \label{HRAN}\vspace*{-10pt}
\end{figure}

According to the described clustering approach above, it is needed to divide all the training samples into pieces and regard one day as the training unit, which consists of 144 items. The average, maximum and minimum values of wind speed and temperature are calculated as the clustering features. Then the sum, maximum and minimum of the power generation are also collected for clustering. For the forecast period, the power generation before this period is regarded as having the same features of power generation with the forecasting period. Then, the data are clustered by the K-means clustering.

Fig. 6 shows the clustering result, and three lines mean that the initial samples are classified into 3 categories. It can be seen that the three categories are much different in wind speed, temperature and wind power, which are the main factors of prediction. In this process, the blue line is represented by $k = 0$, the green line is for the case of $k = 1$, and the red line is for $k = 2$. The initial samples have different features, and the K-means method can classify the samples which have similar features into one category.

\begin{figure}
\centering \vspace*{0pt}
\includegraphics[scale=0.15]{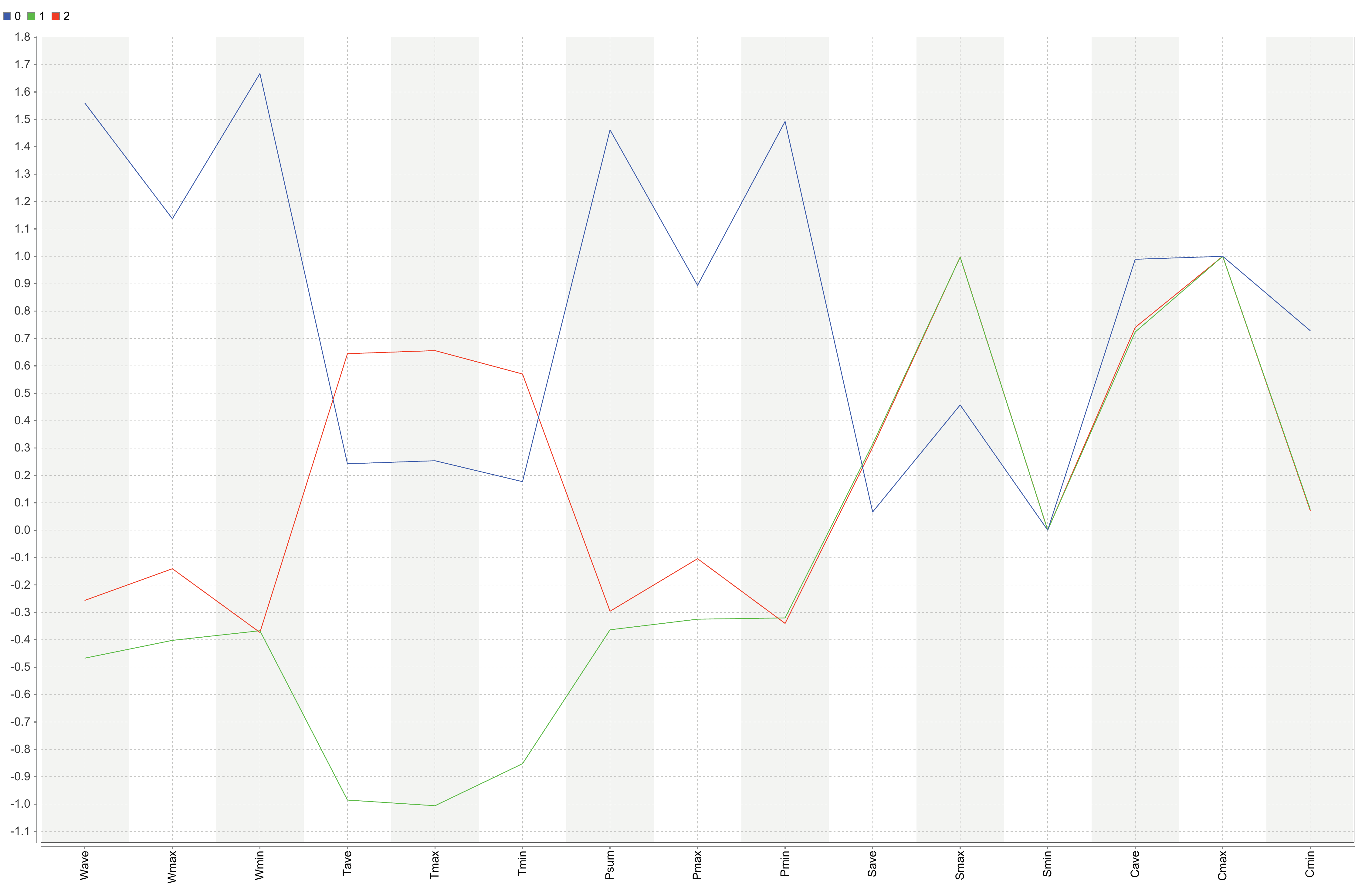}
\setlength{\belowcaptionskip}{-100pt} \caption{\textbf{The K-means clustering results with meteorological conditions and historical power}} \label{HRAN}\vspace*{-10pt}
\end{figure}

The distance between the forecasting day and the 3 categories is calculated according to the Pearson correlation coefficient:
\begin{equation}
P = \frac{{E\left( {XY} \right) - E\left( X \right)E\left( Y \right)}}{{\sqrt {E\left( {{X^2}} \right) - {E^2}\left( X \right)} \sqrt {E\left( {{Y^2}} \right) - {E^2}\left( Y \right)} }}
\end{equation}
where X-axis and Y-axis represent the forecasting sample and the 3 categories, respectively. The most related category is chosen and there are 29 days totally adopted as the training data. In order to compare the performance of clustering with non-clustering, a data set containing 29 days from September 1th to September 29th is used as the training set.

After the feature selection based on Relief algorithm, the training data consisting of the wind speed, the blade angle, and the temperature, are used as the input variables to construct a BPNN model based on the bagging algorithm.

\subsection{Forecasting results}

To better verify the effectiveness of the proposed approach, this paper compares three methods of predicting power generation: the BPNN without clustering pretreatment, the BPNN with clustering pretreatment and the proposed approach.
\begin{figure}
\centering \vspace*{0pt}
\includegraphics[scale=0.25]{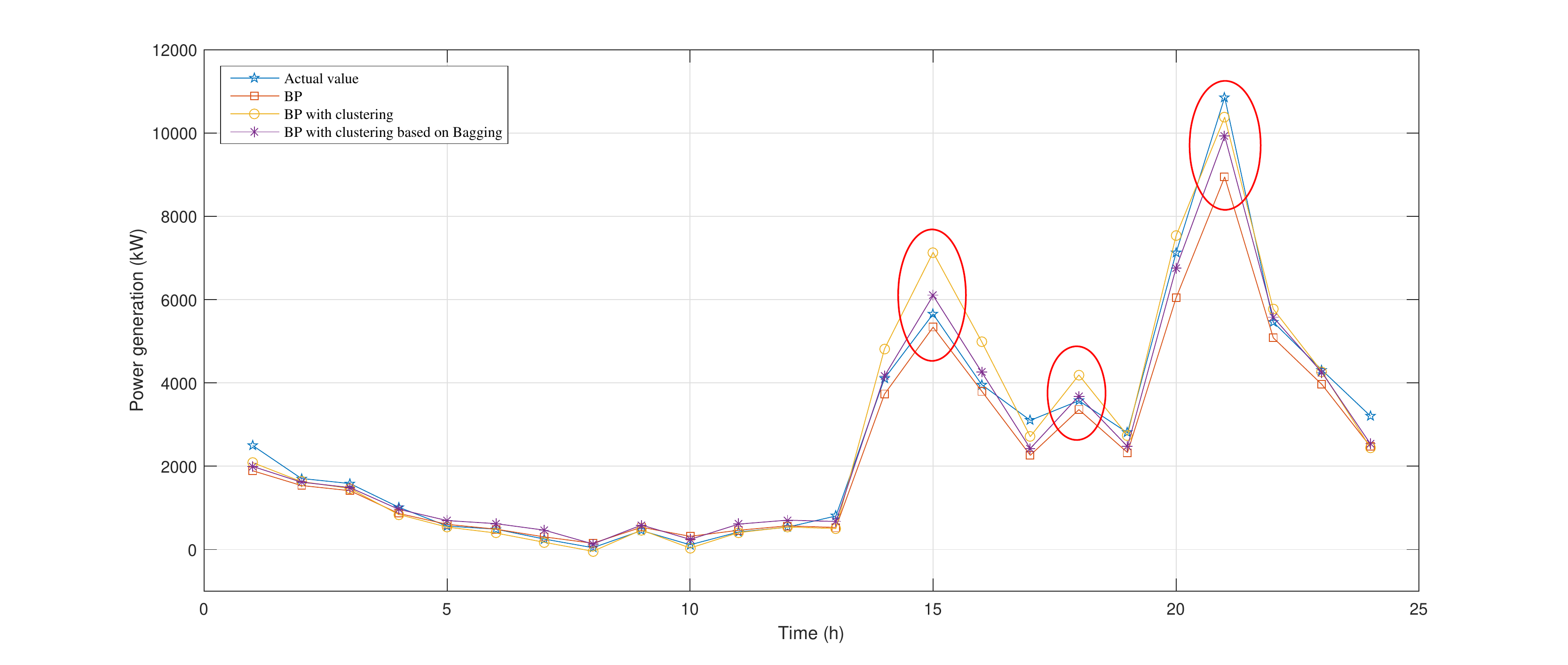}
\setlength{\belowcaptionskip}{-100pt} \caption{\textbf{The 24-hours forecasting results of BPNN, BPNN with clustering and the proposed approach}} \label{HRAN}\vspace*{-10pt}
\end{figure}

Fig. 7 shows the comparison of the three approaches. It is obvious that the purple line is most similar to the actual wind power value. In particular, the peak values marked on Fig. 7 shows that the proposed approach has better performance compared with other approaches.

\begin{table}
\caption{The performance of baseline approaches and the proposed approach}\label{survey}
\scriptsize 
\centering
\begin{tabular}{|p{1.3 in}|p{0.7 in}|p{0.7 in}|}
 \hline
 \textbf{Forecasting approach} & \textbf{RMSE(kW)}& \textbf{MAE(kW)} \\ \hline
 \multirow{1}{2 in}{BP neural network}
 & 558.098 & 365.091 \\ \hline
 \multirow{1}{2 in}{BPNN with clustering}
 & 487.718 & 323.596 \\ \hline
 \multirow{1}{2 in}{Bagging-BPNN with clustering}
 & 342.548 & 255.156\\ \hline
\end{tabular}
\end{table}
From the prediction results in Table \uppercase\expandafter{\romannumeral2}, it can be seen that the BPNN with clustering can reduce the forecasting error compared with the BPNN, in which RMSE and MAE are decreased by 12.7\% and 11.5\%, respectively. Furthermore, the proposal can obtain the best performance among these three approaches, which can decrease 38.7\% and 29.8\% for RMSE compared with the BPNN, and 30.1\% and 21.1\% for MAE compared with the BPNN with clustering. The simulation results verify that the bagging neural network with the similar meteorological conditions and historical power clustering approach can obtain better forecasting accuracy.

Besides, the Approach 1 in \cite{bib:17}, the Approach 2 in \cite{bib:22}, and the Approach 3 in \cite{bib:23} have been simulated and the results are listed in Table \uppercase\expandafter{\romannumeral3}.
\begin{table}
\caption{The performance of three methods and the proposed method}\label{survey}
\scriptsize 
\centering
\begin{tabular}{|p{1.3 in}|p{0.7 in}|p{0.7 in}|}
 \hline
 \textbf{Forecasting approach} & \textbf{RMSE (kW)}& \textbf{MAE (kW)} \\ \hline
 \multirow{1}{2 in}{Approach 1}
 & 1053.736 & 956.211 \\ \hline
 \multirow{1}{2 in}{Approach 2}
 & 427.912 & 307.379 \\ \hline
 \multirow{1}{2 in}{Approach 3}
 & 389.472 & 279.528\\ \hline
 \multirow{1}{2 in}{Bagging-BPNN with clustering}
 & 342.548 & 255.156\\ \hline
\end{tabular}
\end{table}
It can be seen that the proposed approach has the best forecasting accuracy among these approaches. The Approach 1 has the worst results because the different data has different characteristics, and it may be not applicable to this case. The Approach 2 uses the ANN ensemble based on PCA to predict the wind power. The proposal decreases 19.9\% and 17.1\% for RMSE and MAE compared with the Approach 2, respectively. The Approach 3 uses the artificial neural network as the forecasting approach and it has better performances than the Approach 1 and 2, but is worse than the proposal. As a result, the proposal can obtain best forecasting accuracy than the other baselines.

\section{Conclusion}

In this paper, a data mining approach for wind power forecasting has been proposed, which consists of the K-means clustering method and bagging neural network. The historical data are clustered according to the meteorological conditions and historical power. Pearson correlation coefficient is used to calculate the distance between the forecasting day and the clusters. The input variables of the neural network are selected by Relief algorithm to reduce the complexity and the Bagging algorithm is applied to optimize the stability and accuracy of the BPNN. To demonstrate the effectiveness, the proposed approach has been tested according to the actual data in the practical wind farm. The RMSE and MAE results show that the proposals have significant gains.

Although the proposals are not specially designed for the individual wind turbine, the idea of clustering is still important and effective when a large-scale wind farm is built. Particularly, in a wind farm, the location of wind turbines may lie in one direction, then the wind speed of these wind turbines can be classified into one category. With this way, the proposal can be extended and widely used in all real wind farm, which not only increases the forecasting accuracy, but also reduces the computational complexity.

To improve the forecasting accuracy, the effective meteorological forecasting should be researched, and the corresponding optimal method for the BPNN should be designed. Besides, the multi-dimensional clustering problem should be formed and the wind power forecasting model for wind farms should be researched.

\appendices

\begin{IEEEbiography}[{\includegraphics[height=1.2in]{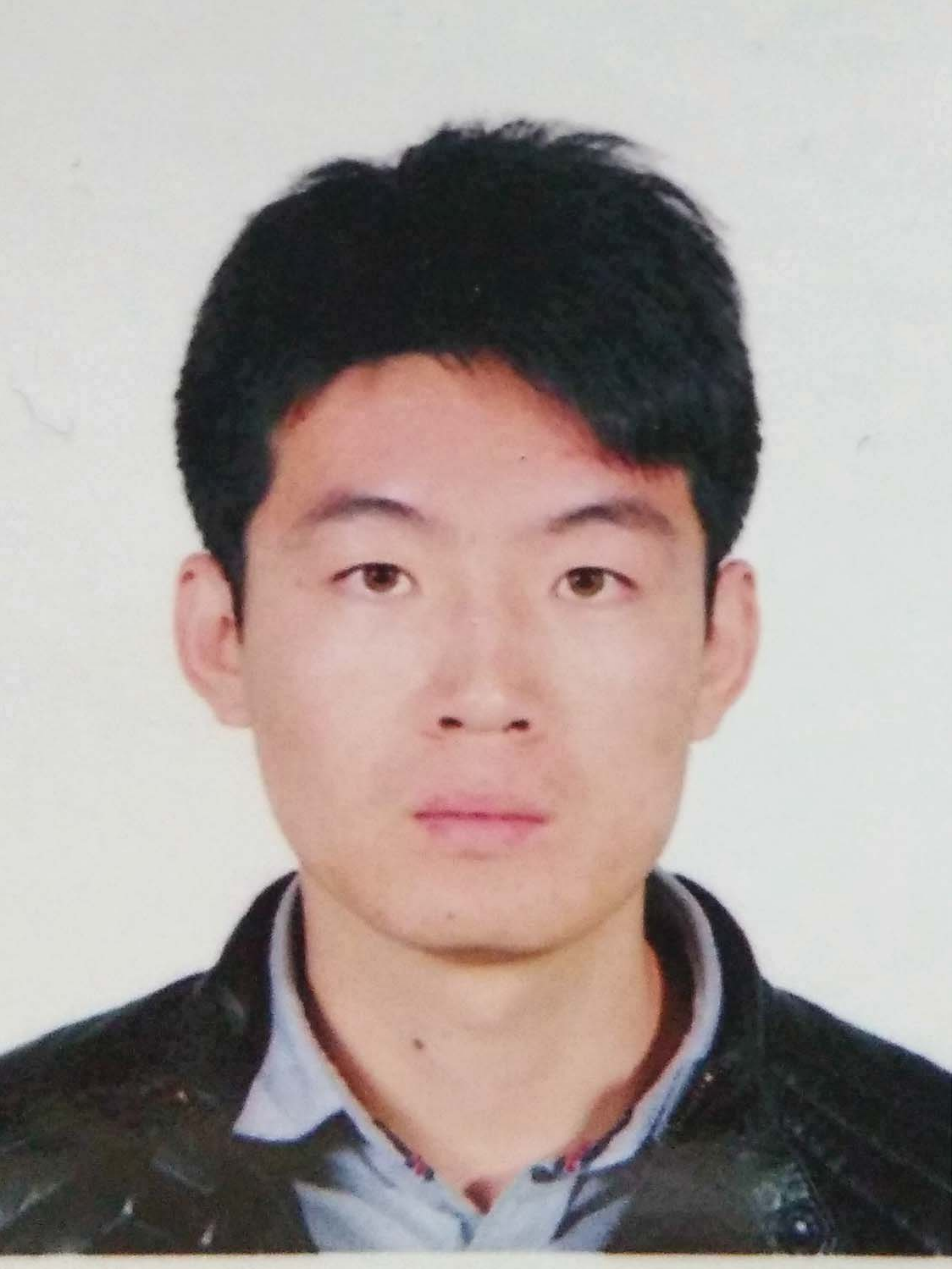}}]{Wenbin Wu}
received his bachelor degree in Telecommunications Engineering with Management from Beijing University of Posts and Telecommunications (BUPT), Beijing, China, in 2014. He is currently a Ph.D. student in the Key Laboratory of Universal Wireless Communications (Ministry of Education) at BUPT. His main research interests include wind power forecasting and non-orthogonal multiple access for cloud radio access network.
\end{IEEEbiography}

\begin{IEEEbiography}[{\includegraphics[height=1.35in]{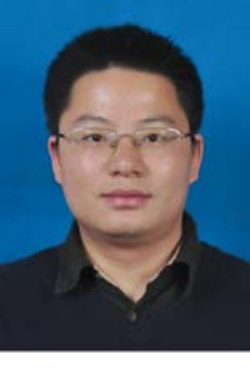}}]
{Mugen Peng} (M'05-SM'11) received the B.E. degree
in electronics engineering from the Nanjing University
of Posts and Telecommunications, Nanjing,
China, in 2000, and the Ph.D. degree in communication
and information systems from the Beijing
University of Posts and Telecommunications (BUPT),
Beijing, China, in 2005. Afterward, he joined
BUPT, where he has been a Full Professor with
the School of Information and Communication Engineering
since 2012. In 2014, he was an Academic
Visiting Fellow with Princeton University, Princeton,
NJ, USA. He leads a Research Group focusing on wireless transmission
and networking technologies with the Key Laboratory of Universal Wireless
Communications (Ministry of Education), BUPT. His main research areas
include wireless communication theory, radio signal processing, and convex
optimizations, with a particular interests in cooperative communication, selforganization
networking, heterogeneous networking, cloud communication,
and internet of things. He has authored/coauthored over 60 refereed IEEE
journal papers and over 200 conference proceeding papers.

Dr. Peng was a recipient of the 2014 IEEE ComSoc AP Outstanding
Young Researcher Award, and the best paper award in IEEE WCNC 2015,
WASA 2015, GameNets 2014, IEEE CIT 2014, ICCTA 2011, IC-BNMT
2010, and IET CCWMC 2009. He received the First Grade Award of the
Technological Invention Award in the Ministry of Education of China, and
the First Grade Award of Technological Invention Award from the China
Institute of Communications. He is on the Editorial/Associate Editorial Board
of the \emph{IEEE Communications Magazine}, \emph{IEEE Access}, \emph{IEEE Internet of
Things Journal}, \emph{IET Communications}, and \emph{China Communications}. He has
been the guest leading editor for the special issues in the \emph{IEEE Wireless
Communications}. He is the Fellow of IET.
\end{IEEEbiography}

\end{document}